\documentclass[aps,preprint,superscriptaddress,showpacs]{revtex4}
\usepackage{graphicx}
\usepackage{dcolumn}
\usepackage{amsmath}

\begin{document}
\bibliographystyle{apsrev}

\title{Dynamically Coupled Oscillators 
-- Cooperative Behavior via Dynamical Interaction --
}
 
\author{Toru Aonishi}

\affiliation{Brain Science Institute, RIKEN, 
2-1 Hirosawa, Wako-shi, Saitama, 351-0198, Japan 
}
\author{Masato Okada}

\affiliation{Brain Science Institute, RIKEN, 
2-1 Hirosawa, Wako-shi, Saitama, 351-0198, Japan 
}

\date{\today}

\begin{abstract}
We propose a theoretical framework to study the cooperative behavior of
dynamically coupled oscillators (DCOs) that possess dynamical
interactions. Then, to understand synchronization phenomena in networks
of interneurons which possess inhibitory interactions, we propose a DCO
model with dynamics of interactions that tend to cause 180-degree phase
lags. Employing an approach developed here, we demonstrate that although
our model displays synchronization at high frequencies, it does not
exhibit synchronization at low frequencies because this dynamical
interaction does not cause a phase lag sufficiently large to cancel the
effect of the inhibition. We interpret the disappearance of
synchronization in our model with decreasing frequency as describing the
breakdown of synchronization in the interneuron network of the CA1 area
below the critical frequency of 20 Hz.
\end{abstract}


\maketitle

Many studies of cooperative behavior in many-body systems, of which one
prototype is represented by spin systems, have been based on the
assumption that the dynamics of interactions can be ignored.  However,
in the modeling of many biological systems, the inclusion of interaction
dynamics is necessary to obtain a faithful description. In much of the
research that has been done, systems with such {\it slow} interaction
have been mimicked by a so-called {\it multi-compartment model} which
corresponds to a discrete version of a {\it distributed parameter
system}. By using such a multi-compartment model, we can accurately
estimate the dynamical behavior of a single unit. However, it is
impossible to carry out a large-scale numerical simulation with such
models to understand the behavior of a large-scale system composed of
tens of thousands of units. Even if it is possible, we cannot understand
the essence of such phenomena. To address this problem, we need to
reduce the  degrees of freedom needed in a descriptive model.

Phase reduction is a powerful tool that helps us understand the
oscillatory behavior of large-scale systems composed of such
multi-compartment models. In previous studies of systems consisting of
such multi-compartment models, phase descriptions of the entire systems
including the dynamical interactions themselves were obtained
numerically \cite{hansel,vreeswijk,crook1,crook2}.  Thus, the effects of
the interaction dynamics were numerically incorporated into only a {\it
phase-coupling function} to establish the phase equation.  Therefore, in
general, the effects of the interaction dynamics alone cannot be
extracted in such methods.

In this paper, we analyze systems of dynamically coupled oscillators
(DCOs), and demonstrate how the dynamics of the interactions influence
the cooperative behavior exhibited by the systems. Using the multi-scale
perturbation method (MSPM) \cite{ermentrout,kuramoto0} to carry out a
phase reduction, we are able to represent the nonlinear dynamical
interaction in terms of only the amplitude and phase of the fundamental
frequency component in the frequency response of the interaction.  In
control engineering, the amplitude and phase of the fundamental
frequency component of the frequency response are referred to as the
{\it describing function}.  The describing function used in the
treatment of non-linear systems is an extension of the transfer function
used in the treatment of linear systems.  In contrast to previous work
\cite{hansel,vreeswijk,crook1,crook2}, the effects of the interaction
dynamics are expressed by the describing function in our approach, and
so it is easy to isolate and extract the effects of the interaction
dynamics from the phase equation. Therefore, our approach can clearly
elucidate  the effect of the interaction dynamics on the cooperative
behavior of entire systems.  Our results imply the possibility of a new
modeling approach for biological systems by which the describing
function of the interaction is identified. The description proposed here
corresponds more to higher order approximation than to conventional
linear description (e.g., the auto-regressive (AR) model).

In the first half of this paper, we show in detail the process of
analytical phase reduction for a general model with dynamical
interaction. We treat a large population of Stuart-Landau (SL)
oscillators coupled through nonlinear dynamical interaction. The SL
oscillators have the essential structure of the Hopf-bifurcation,
because its evolution equation can be derived from any non-linear
oscillator system with the Hopf-bifurcation through perturbation
expansion \cite{kuramoto0}. In the latter half of the paper, we apply
our theoretical framework to the analysis of a network of interneurons.
The analysis of an interneuron network is a good example to verify the
applicability of our approach for understanding the neural cooperative
behavior via dynamical interaction.

First, we consider a large population of SL oscillators
weakly coupled through interactions which themselves are 
nonlinear dynamical systems.
We analyze the following coupled system:
\begin{eqnarray}
\hspace{-0.2cm}
\frac{d w_j}{d t} = w_j \left(1-|w_j|^2\right) 
+ i \left(\Omega + \epsilon \omega_j \right) w_j +  
\frac{\epsilon}{N} \sum_{k=1}^N s_{j k}. \label{eq.mod1} 
\end{eqnarray}
Here $w_j$ is the state variable (a complex number) 
of the $j$th oscillator (with a total of $N$). 
$\Omega$ is the average natural frequency and
$\epsilon \omega_j$ represents the deviation from the average 
for the $j$th oscillator randomly distributed 
with a density represented by $g(\omega)$.
The terms multiplied by $\epsilon$ are considered perturbations,
and thus, the quantity $\epsilon$ controls the magnitude of 
perturbations. When $\epsilon=0$, the system 
has an unstable fixed point at the origin
and a stable limit-cycle orbit on the unit circle in the complex plane
represented by  
\begin{eqnarray}
w_j(t) = \Phi_j(t), \ 
\Phi_j(t) = \exp i \left( \Omega t + \phi_j \right),\ 
^\forall \phi_j \in {\bf R}/2\pi,
 \label{eq.orbit}
\end{eqnarray}
where $\phi_j$ is the phase of the $j$th oscillator.
When $\epsilon = 0$, the system is neutrally stable with respect
to a perturbation in the form of a temporal shift
while it conserves a fixed orbit.
Thus, $\phi_j$ depends only on the initial condition. 
The quantity $s_{j k}$ in Eq. (\ref{eq.mod1}) 
represents the output from the dynamical system 
of the interaction expressed by 
\begin{eqnarray}
s_{j k} = V_{j k} (x_{j k}, w_{k}), \ \ \frac{d x_{j k}}{d t} &=& F_{j k} \left(x_{j k}, w_{k} \right),
\label{eq.int_g}
\end{eqnarray}
where $x_{j k}$ represents the internal state of the interaction, 
$F_{jk}(\cdot)$ is a nonlinear function 
determining its dynamics, and $V_{jk}(\cdot)$ is an output function. 

We now discuss the dynamical interaction described by Eq. (\ref{eq.int_g}).
If $w_k = \Phi_k(t) = \exp i \left( \Omega t + \phi_k \right)$
is input into Eq. (\ref{eq.int_g}), 
the higher harmonics resulting from the nonlinearity of 
this equation are superimposed on $s_{j k}$.
Here we restrict our consideration to the case that the output $s_{j k}$
is a periodic function possessing the same period as the input $w_k$.
In this case, $s_{j k}$ can be expanded into the following Fourier form: 
$s_{j k} = J_{j k}^{(1)} \Phi_k + \sum_{n>1}^\infty J_{j k}^{(n)} \Phi_k^n$.
If Eq. (\ref{eq.int_g}) takes a form of a linear system,
$J_{j k}^{(1)}$ would be non-zero, and 
$J_{j k}^{(n)}$ for $n>1$ would all be zero.
In this case, $J_{j k}^{(1)}$ is called the {\it transfer function}
of the dynamical interaction Eq. (\ref{eq.int_g}).
If, on the other hand, Eq. (\ref{eq.int_g}) 
takes a form of a nonlinear system,
$J_{j k}^{(1)}$ and some of  $J_{j k}^{(n)}$ for $n>1$ 
would be non-zero. In this case, $J_{j k}^{(1)}$ is called the
{\it describing function} in the field of control engineering. 

Now, employing the MSPM, we attempt to 
reduce Eq. (\ref{eq.mod1}) to a phase equation.
The solution of the perturbed system (\ref{eq.mod1}) can be 
represented as 
\begin{eqnarray}
w_j(t) = \Phi_j(t) + \epsilon u_j(t),\  \Phi_j(t) = \exp i \left( \Omega t + \phi_j(\epsilon t)\right), \label{eq.sol1} 
\end{eqnarray}
where we write $\phi_j$ as a function of $\epsilon t$
to make explicit its slow evolution in time, 
and $\epsilon u_j$ is the first order change in $w_j$ introduced by the
perturbation. If we substitute Eq. (\ref{eq.sol1}) into
Eqs. (\ref{eq.mod1}) and (\ref{eq.int_g}) and expand the resulting 
expression about $\epsilon=0$, the $O(\epsilon)$ equation becomes
\begin{eqnarray}
i \Phi_j \frac{d \phi_j}{d (\epsilon t)} &=&  i \omega_j \Phi_j 
+ {\it L}_{\phi_j} u_j + \frac{1}{N} \sum_{n=1}^\infty \sum_{k=1}^N J_{j k}^{(n)} \Phi_k^n. 
\label{eq.exp1}
\end{eqnarray}
Here ${\it L}_{\phi}$ is the linear operator 
corresponding to Eq. (\ref{eq.mod1}) linearized about 
the periodic solution (\ref{eq.orbit}) in the case $\epsilon=0$.
It is given by  
\begin{eqnarray}
{\it L}_{\phi} u = - \frac{du}{dt} + i \Omega u - \Phi^2 \overline{u} -
u, 
\end{eqnarray}
where the overline denotes complex conjugation. All of the 
eigenvalues of ${\it L}_{\phi}$ are non-positive,
since the solution $\Phi(t)$ is stable.
Note that there exists the eigenfunction $\Phi'(t)$ of 
${\it L}_{\phi}$ with eigenvalue $0$. 
This eigenfunction corresponds to an infinitesimal temporal shift because 
$\Phi(t+ \delta) \sim \Phi(t) + \delta \Phi'(t)$.
We assume there exists no other eigenfunction with eigenvalue $0$ 
in the space of the periodic functions, so that 
${\rm ker} {\it L}_{\phi} = {\rm span}\{ \Phi'(t)\}
= {\rm span}\{ i \Phi(t)\}$.
This assumption is equivalent to that of the orbital stability of 
$\Phi(t)$.

We define the inner product of two $\frac{2\pi}{\Omega}$-periodic
complex functions $u(t)$ and $v(t)$ as
$\left< u(t), v(t) \right> \nonumber =\int_{0}^{\frac{2\pi}{\Omega}} dt 
\left({\rm Re}\{u(t)\} {\rm Re}\{v(t)\} + {\rm Im}\{u(t)\} {\rm Im}\{v(t)\}
\right)$.
Then, we obtain the operator adjoint to $L_{\phi}$ as  
\begin{eqnarray}
{\it L}_{\phi}^\ast u = \frac{du}{dt} - i \Omega u - \Phi^2 \overline{u} - u.
\end{eqnarray}
From Fredholm's alternative \cite{ermentrout}, there is 
a so-called {\it response function} $\Phi^\ast(t)$ that
spans the kernel of $L^{\ast}_\phi$ in the space of periodic functions. 
Therefore, ${\rm ker} {\it L}_{\phi}^\ast = {\rm span}\{\Phi^\ast(t)\}$.
In this case, we explicitly obtain $\Phi^\ast(t) = i \Phi(t)$.
Taking the inner product of $i \Phi(t)$ and Eq. (\ref{eq.exp1}), we obtain
\begin{eqnarray}
& &\left< i \Phi_j,  i \Phi_j \right> \frac{d \phi_j}{d (\epsilon t)} - \left< i \Phi_j,  i \Phi_j \right> \omega_j - \frac{1}{N} \sum_{k, n} \left< i \Phi_j, J_{j k}^{(n)} \Phi_k^n \right> \nonumber \\
& &=\left< i \Phi_j,  {\it L}_{\phi_j} u_j \right> = \left<{\it L}_{\phi_j}^\ast i \Phi_j,  u_j \right> = 0,
\end{eqnarray}
where the quantity $\epsilon t$ is treated as a constant over a single
period, since $\phi_j(\epsilon t)$ is driven 
by a weak perturbation. Here, the higher harmonics  
resulting from the nonlinearity of Eq. (\ref{eq.int_g}) cancel, 
because $\left< i \Phi_j, J_{j k}^{(n)} \Phi_k^n \right>=0\ 
(n>1)$. Thus, we derive the phase
equation describing the slow phase dynamics of the system,
\begin{eqnarray}
& &\frac{d \phi_j}{d (\epsilon t)} 
= \omega_j + \frac{1}{N} \sum_j^N A_{j k}
\sin\left(\phi_k - \phi_j + \psi_{j k} \right),
\label{eq.phase}\\
& &{\rm with}\ A_{j k} = |J_{j k}^{(1)}|, \ \ \psi_{j k} = \arg J_{j k}^{(1)}, 
\label{eq.tra1}
\end{eqnarray}
where $\omega_j$ represents a natural frequency of unit $j$ at the level of 
the phase equation. As Eqs. (\ref{eq.phase}) and (\ref{eq.tra1}) reveal,  
in the phase equation, the dynamical interaction (\ref{eq.int_g}) 
is expressed in terms of the describing function. 
Thus, seen the fundamental frequency components in the output from 
the dynamical system of the interaction (\ref{eq.int_g}) clearly play a key role 
in the slow phase dynamics of the system.

If all interactions are identical, i.e. $J_{j k}^{(1)}=J$ 
($A_{j k} = A$ and $\psi_{j k} = \psi$), Eq. (\ref{eq.phase})
is equivalent to the Sakaguchi-Kuramoto model \cite{sakaguchi},
and so, in this case, the Sakaguchi-Kuramoto theory 
can be applied to the analysis of Eq. (\ref{eq.phase}). 
Here, we define the mean field as 
$m = \frac{1}{N} \sum_{j=1}^N w_j$. 
When $|m| \neq 0$, the system is in a state of phase
synchronization. In the thermodynamic limit, $N\rightarrow \infty$,
we obtain the following self-consistent equation relating the order parameters $|m|$
and $\tilde{\Omega}$: 
\begin{eqnarray}
|m| e^{i \psi} &=& A |m| \int_{-\pi/2}^{\pi/2} d \phi g\left(\tilde{\Omega} +
|m|\sin\phi \right)\cos\phi\exp(i \phi)\nonumber \\ 
&+& i  A |m| \int_{0}^{\pi/2} d \phi 
\frac{\cos\phi (1-\cos\phi)}{\sin^3\phi} \nonumber \\ 
& &\times \left\{ 
g\left(\tilde{\Omega}+\frac{|m|}{\sin\phi}\right)-g\left(\tilde{\Omega}-\frac{|m|}{\sin\phi}\right)
\right\}. \label{Eq.X}
\end{eqnarray}
Here, to derive the order parameter equation, we have assumed the
existence of one large cluster of oscillators 
phase-locked at frequency $\tilde{\Omega}$ in the slow dynamics described by 
(\ref{eq.phase}). In the original system, described by
(\ref{eq.mod1}), the frequency of the synchronous cluster is $\Omega+\epsilon \tilde{\Omega}$.

\begin{figure}
(a)\includegraphics[height=4.5cm]{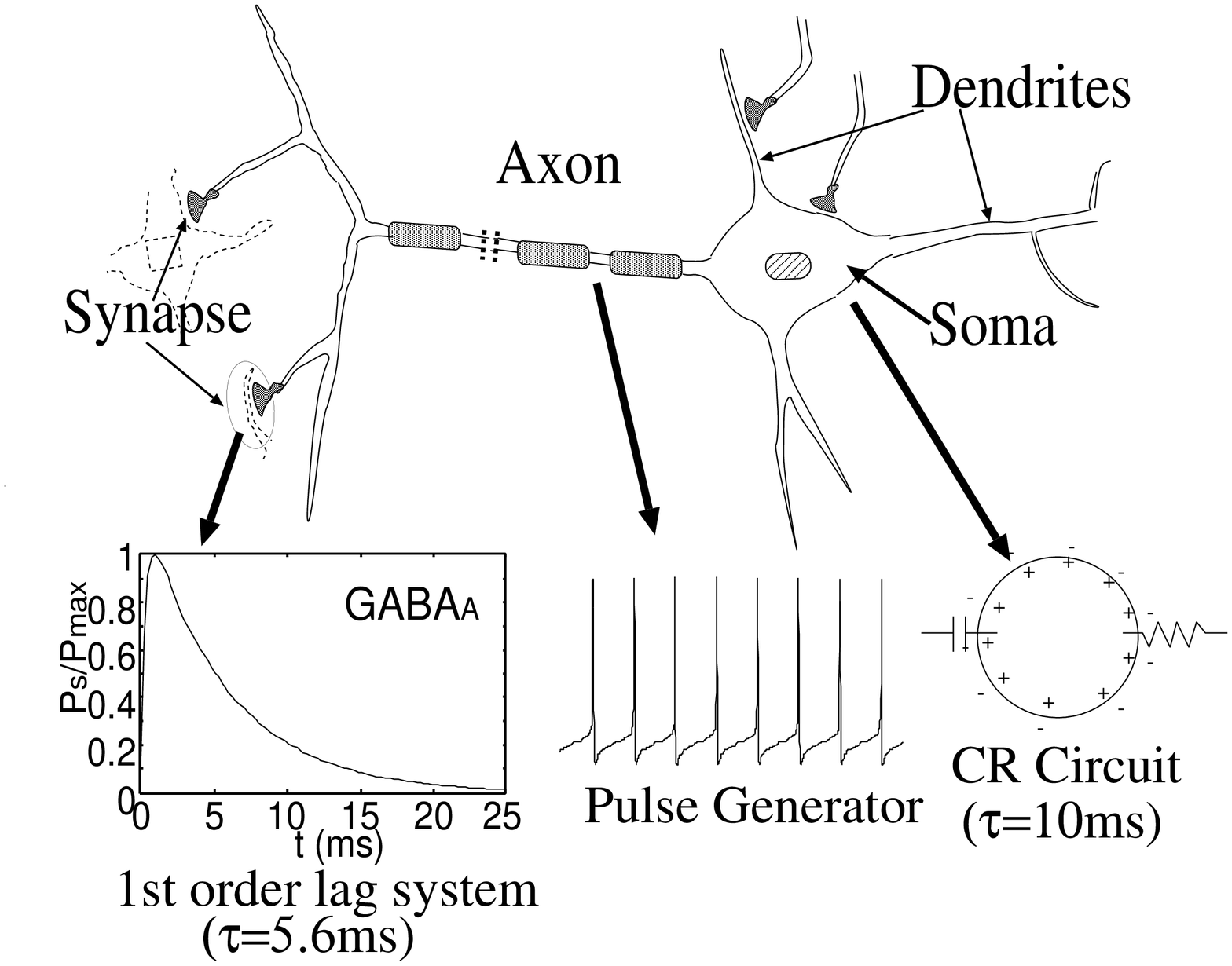}\\
(b)\hspace{-0.5cm}\includegraphics[width=4.7cm]{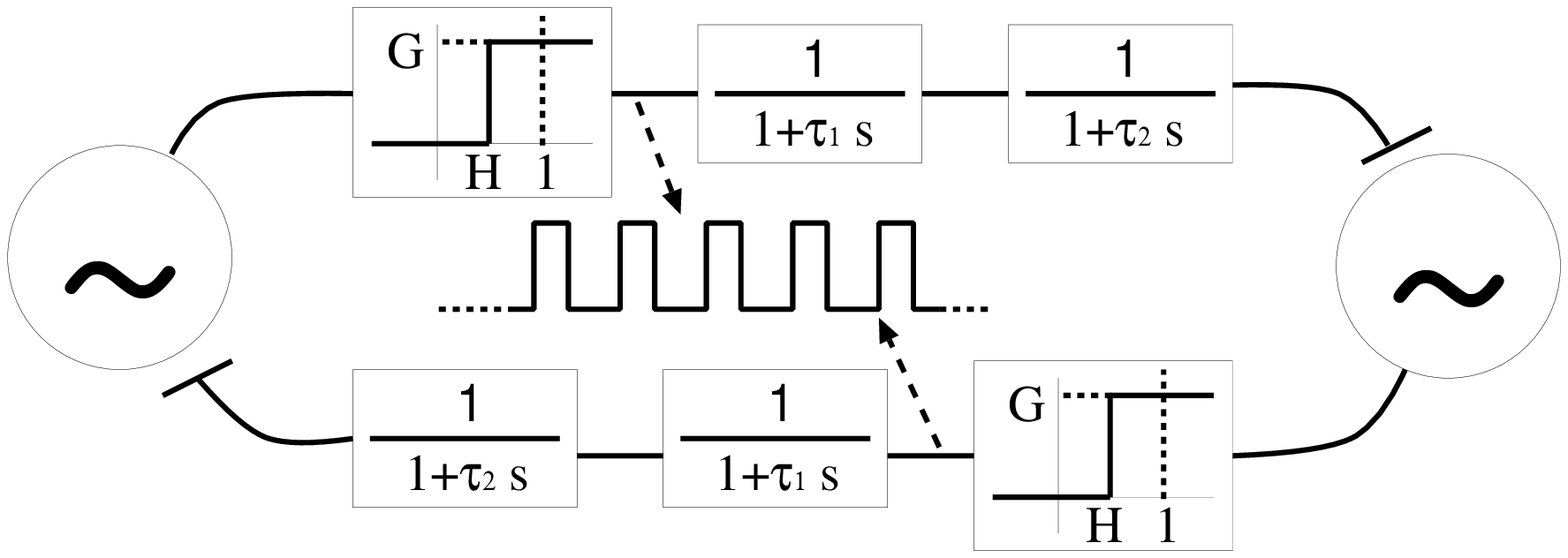}
(c)\hspace{-0.5cm}\includegraphics[width=4.0cm]{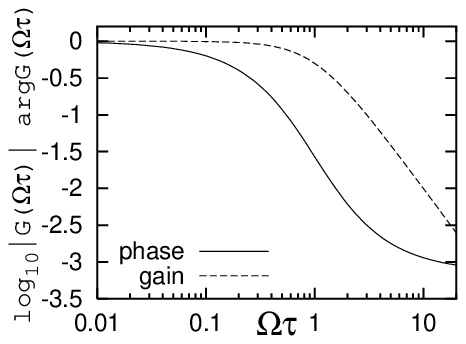}
\caption{(a) Dynamical behavior of neural components.
(b) Schematic diagram of the dynamical interaction employed 
in our model which consists of a rectangular-pulse generator and 
a second-order lag system (SOLS). (c) Bode diagram of 
the SOLS, $G(i\Omega)=1/(1+i\Omega\tau)^2$.}
\label{model}
\end{figure}

Next, we apply our theoretical framework
to the analysis of a network of interneurons as an example.
High frequency oscillations, known as 
$\gamma$ oscillations, have been observed in many areas of the brain.
The synchronization of inhibitory interneurons  
is known to be the origin of $\gamma$ oscillations \cite{Jef1}. 
Inhibitory interactions, which correspond to 
anti-ferromagnetic interactions in spin systems, 
introduce {\it competition}; for this reason, in general, 
such interactions tend to prevent synchronization.
Therefore, the synchronization phenomena in networks of interneurons 
are non-trivial \cite{vreeswijk,udo,wang}.
A series of physiological {\it in vitro} experiments  
inspired our study. One in particular was the experimental discovery that 
the interneuronal network in the CA1 area of the hippocampus 
displays synchronization at
frequencies greater than or close to 20 Hz, but not at
frequencies significantly below this value \cite{Jef2,Jef3}. 
The discovery of this phenomenon provides evidence suggesting that 
cooperative behavior in networks of interneurons arises due to the
dynamical nature of the interactions.

To understand the mechanism of the synchronization, 
we propose a DCO model with interaction dynamics that tend to cause 180-degree phase lags. 
The estimated connectivity between interneurons in the CA1 area is 10\%
\cite{wang}. The interneuron network in the CA1 area has dense connectivity,
because the number of connections between interneurons is $O(N)$.
Therefore, we can capture properties of the interneuron network
by analyzing the global coupling systems.
As shown in Fig. \ref{model}(b), the dynamical interactions of
neurons are modeled by a rectangular-pulse generator consisting of  
a threshold element and a second-order lag system (SOLS):
\begin{eqnarray}
& &\hspace{-0.8cm}s_{j k} = -x_{j k}^{(1)},\ 
\left[ \begin{array}{c}
\tau^{(1)}_{j k} \frac{d x_{j k}^{(1)}}{dt}\\
\tau^{(2)}_{j k} \frac{d x_{j k}^{(2)}}{dt}
\end{array}
\right]
= \left[ \begin{array}{cc}
-1 & 1\\
0 & -1
\end{array}
\right]
\left[ \begin{array}{c}
x_{j k}^{(1)}\\
x_{j k}^{(2)}
\end{array}
\right] \nonumber\\
& &\hspace{-0.4cm}+ \left[ \begin{array}{c}
0\\
G
\end{array}
\right]
\Theta\left({\rm Re}\{w_k\} - H \right), \ \Theta(x) = \left\{
\begin{array}{cc}
1 & x > 0 \\
0 & x\leq 0
\end{array}
\right.. \label{eq.int1} 
\end{eqnarray}
The parameters $\tau^{(1)}_{j k}$ and $\tau^{(2)}_{j k}$ are 
the time constants of the two cascade units of the first-order 
lag system that form the SOLS. 
In Eq. (\ref{eq.int1}), the parameters $G$ and $H$ 
denote, respectively, the gain and the threshold of the threshold element,
and are defined as $G = 1/\theta$ and $H = \cos \theta$ 
$(\theta>0)$. If the value of $\theta$ is small, 
the rectangular pulse generated by the threshold element is sharp. 
Note that Eq. (\ref{eq.int1}) represents   
mutual interactions only between the real parts of the SL oscillators.

In the context of control engineering, 
the SOLS is regarded as one of the minimal models for 
realizing a 180-degree phase lag. 
As shown in Fig. \ref{model}(a), the electrical properties 
of somata and proximal dendrites can be approximated as the CR circuit.
Near resting potential, we can obtain a linear approximation 
of their behavior as being that of a first-order lag system. 
The response of the synaptic conductance is also 
displayed in Fig. \ref{model}(a). 
The evolution of the synaptic conductance
is roughly reproduced by that of the first-order lag system.
Thus, as shown in Fig. \ref{model}(a), the SOLS given by Eq. (\ref{eq.int1}) 
can be considered as a cascade system consisting of these components. 
In addition, to model the generation of action 
potentials in a neuron when its membrane 
potential reaches the threshold, and furthermore, 
to demonstrate the applicability of our approach in analyses of 
non-linear systems, we have also introduced a threshold element into 
the dynamical interaction given in Eq. (\ref{eq.int1}). 

The describing function for the dynamical interaction given in 
(\ref{eq.int1}) is
\begin{eqnarray}
J_{j k}^{(1)} &=& - \frac{2 \sin \theta}{\pi \theta} \cdot \frac{1}{(1+i \tau^{(1)}_{j k} \Omega)(1+i \tau^{(2)}_{j k} \Omega)}. \label{eq.tra2}
\end{eqnarray}
Here, the describing function of the threshold element is  
$2 \sin \theta/(\pi \theta)$, which contributes  only to 
the gain of Eq. (\ref{eq.tra2}). 
Figure \ref{model}(c) displays the Bode diagrams 
of the SOLS. As Fig. \ref{model}(c) reveals, the gain of the SOLS,
which is a type of low-pass filter, 
is a decreasing function of frequency at high frequencies,
while the phase of the SOLS converges to -180 degrees
as the frequency increases. 
If the phase lag is 180 degrees, $J_{j k}^{(1)}$ is a positive real number, 
and the system thus effectively becomes a ferromagnetic 
oscillator system. Note that Eq. (\ref{eq.int1}) represents 
the mutual interactions only between the real parts of the 
SL oscillators. Thus, $A_{j k}$ takes half the value it had 
in Eq. (\ref{eq.tra1}), and here $A_{j k} = \frac{1}{2}|J_{j k}^{(1)}|, \ \ 
\psi_{j k} = \arg J_{j k}^{(1)}$. 

\begin{figure}
(a)\hspace{-0.5cm}\includegraphics[height=3cm]{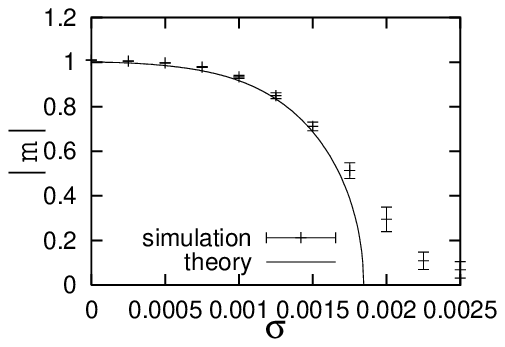}
(b)\hspace{-0.5cm}\includegraphics[height=3cm]{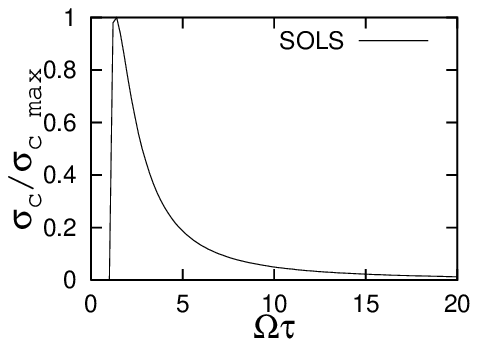}
\caption{(a) $|m|$ as a function of $\sigma$.
The data points with error bars represent the means and standard deviations 
of 10 samples obtained from numerical simulations of Eq. (\ref{eq.mod1}).  
Here, $\Omega=2 \pi 80$, $\tau=0.02$, 
$\theta=\pi/6$, $\epsilon=10$ and $N=1000$.
(b) Critical deviation  $\sigma_c$ as a function of 
$\Omega \tau$. The quantity plotted here is actually a normalized
 critical deviation, with maximum value $1$. 
}
\label{av1}
\end{figure}

In the numerical calculation we will now 
discuss, we chose the distribution of natural frequencies as 
$g(\omega)=(2\pi\sigma^2)^{-1/2} \exp\left(-\omega^2/(2\sigma^2)\right)$,
where $\sigma$ is the deviation of the natural frequencies. 
We begin with systems where the interactions are all equivalent, i.e., 
$\Omega \tau^{(1)}_{j k}=\Omega, \tau^{(2)}_{j k}=\Omega \tau$,
and hence $A_{jk}=A$ and $\psi_{jk}=\psi$. 
In this case, Eq. (\ref{eq.phase})
is equivalent to the Sakaguchi-Kuramoto model \cite{sakaguchi}.
Figure \ref{av1}(a) displays $|m|$ as a function of $\sigma$
in the case of the dynamical interaction given by Eq. (\ref{eq.int1}), 
where the solid curves were obtained from the order parameter equation 
(\ref{Eq.X}), and the data points with error bars represent results 
obtained using the fourth-order Runge-Kutta method with
Eqs. (\ref{eq.mod1}) and (\ref{eq.int1}).
Next, we estimate the critical value of $\sigma$, 
representing the boundary between $|m|=0$ and $|m|\neq 0$.
Figure \ref{av1}(b) displays the critical deviation  
$\sigma_c$ as a function of $\Omega \tau$ for the two models. 
Note that the absolute values of $A$ and $\sigma_c$ have no meaning
in comparing such models, because $\sigma_c$ is proportional to 
$A$ \cite{sakaguchi}. Plotted in Fig. \ref{av1}(b) is the critical 
deviation normalized to have a maximum value of $1$.
The widths of the synchronous regions decrease as the frequency increases
because, as shown in Fig. \ref{model}(c), the SOLS is
a low-pass filter whose gain rapidly decreases at higher frequencies. 
However, if $G$ in Eq. (\ref{eq.int1}) is sufficiently large, 
the gain of the interaction given by Eq. (\ref{eq.int1})
can be made sufficiently large by the pulse
generator to realize synchronization at high frequencies.
However, as Fig. \ref{av1}(b) reveals, our model does not exhibit synchronization at low frequencies
because, as shown in Fig. \ref{model}(c),
this dynamical interaction does not 
cause a phase lag sufficiently large to cancel the effect of the inhibition. 
We interpret the disappearance of synchronization in our model 
with decreasing frequency as describing the breakdown of synchronization in the interneuron network of the CA1 area 
below the critical frequency of 20 Hz \cite{Jef2,Jef3}.
Based on this correspondence, we conjecture that in an interneuron network, 
a phase lag resulting from the dynamical nature of the interactions
might cancel the effect 
of inhibition at high frequencies, and through this cancellation, 
the system effectively becomes a ferromagnetic oscillator system.

\begin{figure}
\includegraphics[height=3.0cm]{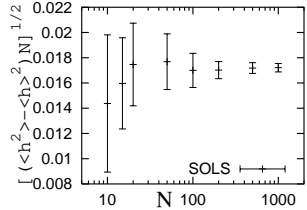}
\caption{System size scaling.
Deviation of local fields multiplied by $\sqrt{N}$ as a function of 
$N$. }
\label{size1}
\end{figure}

By means of numerical simulations, as shown in Fig. \ref{size1}, we have
demonstrated through a scaling plot that the variation of a local field
is $O(1/\sqrt{N})$ in systems where a quenched random time constant is
used for the interaction dynamics in Eq. (\ref{eq.int1}).  In the
thermodynamical limit $N\rightarrow\infty$, the system asymptotically
approaches a system with global homogeneous interaction, even with a
quenched random time constant of the interaction dynamics.  Therefore,
as the system size $N$ increases, the system tends to synchronize
through the cancellation of fluctuation via interaction.

\bibliography{ao_ref.bib}

\end{document}